\documentclass[%
superscriptaddress,
preprint,
amsmath,amssymb,
 ps,
showpacs,
prb,
]{revtex4-2}

\usepackage{graphicx}
\usepackage{dcolumn}
\usepackage{bm}

\begin{document}


\title{Thermally assisted Pauli spin blockade in double quantum dots
}

\author{M. Kondo}
\altaffiliation[Present address:]{Toppan Printing co., ltd.}
\affiliation{College of Engineering, Nihon University, Koriyama, Fukushima, 963-8642, Japan}

\author{S. Miyota}
\affiliation{College of Engineering, Nihon University, Koriyama, Fukushima, 963-8642, Japan}

\author{W. Izumida}
\affiliation{Department of Physics, Tohoku University, Sendai, Miyagi, 980-8578, Japan}

\author{S. Amaha}
\affiliation{RIKEN, Center for Emergent Matter Science, 3-1 Wako, Saitama, 351-0198, Japan}

\author{T. Hatano}%
\email{hatano.tsuyoshi@nihon-u.ac.jp}
\affiliation{College of Engineering, Nihon University, Koriyama, Fukushima, 963-8642, Japan}

\date{\today}

\begin{abstract}
We investigate the influence of thermal energy on the current flow and electron spin states in double quantum dots in series. 
The quadruplet Pauli spin blockade, which is caused by the quadruplet and doublet states, occurs 
at low temperatures affecting the transport properties.  
As the temperature increases, the quadruplet Pauli spin blockade occurs as a result of the thermal energy, even in regions 
where it does not occur  at low temperatures. This is because the triplet state is formed in one dot as a result of the gradual change of the Fermi distribution function of the electrodes with increasing temperature.  
Moreover, the thermally assisted Pauli spin blockade results in coexistence of the Coulomb and Pauli spin blockades.   
Conversely, for the standard triplet Pauli spin blockade,  which occurs as a result of the triplet and singlet states,  
the current through the double dots monotonously smears out
as the temperature increases.  
Therefore, the thermally assisted Pauli spin blockade is not clearly observed. 
However, the coexistence of the Coulomb and triplet Pauli spin blockades as a result of  the thermal energy 
is clearly obtained in the calculation of the probability of  the spin state in the double dots.  
\end{abstract}
\pacs{73.63.Kv, 73.23.Hk}

\maketitle


\section{Introduction}

Quantum information processing via nanoscale devices is increasingly attracting attention. 
Of such nanoscale devices, semiconductor quantum dot devices, which can manipulate electron spins one by one, 
 are expected to be applied to quantum computers \cite{Loss}.  

For double quantum dots in series, 
that are formed by electrostatically and quantum mechanically coupling two quantum dots, 
when an electron is effectively localized in a quantum dot, 
the current through the coupled dots is suppressed by the Pauli spin blockade, 
which forbids transitions between the triplet and singlet states according to the Pauli exclusion principle  \cite{Ono}. 
The conventional Pauli spin blockade with a spin triplet state (the triplet Pauli spin blockade) 
is used for the initialization and readout of a qubit;  
therefore, the triplet Pauli spin blockade is a key technology in the application of quantum dot quantum computers.  
The coherent manipulation of single spins \cite{Koppens,Laird,Nowack,Laird,Pioro} and two-spin entanglement \cite{Petta,Brunner} 
have been reported,  
and more recently, a hybrid structure containing a single-spin qubit and a triplet singlet qubit 
and the high-temperature operation of a silicon qubit have also been reported \cite{Noiri,Ono2}.
For double quantum dots in series, 
the quadruplet state is formed with one conducting and two localized electrons. 
In this case, the tunneling of the conducting electron is forbidden by Pauli spin exclusion 
as a result of the quadruplet state and a quadruplet Pauli spin blockade occurs \cite{Amaha}. 

The Coulomb and Pauli spin blockades are observed with a finite applied source drain voltage. 
This corresponds to a non-equilibrium system;   
therefore, it is necessary but difficult to explain the correlation between the electrons in the ground states and the excited electrons, 
because the current flows not only via the ground state but also via the excited states. 
Further, because the current suppression can be fundamentally explained by a qualitative discussion of the relationship between the spin states 
and their energies \cite{Hanson}, extensive quantitative studies of the relationship between the current suppression and the properties of the spin states are still not completely understood.

Recently, the relationship between thermal energy and quantum mechanical effects has been intensively studied, 
and findings concerning the spin Seebeck effect \cite{Uchida} and a quantum heat engine using a single-spin qubit \cite{Ono3} have been reported. 
Therefore, it is possible to discover new physics with respect to thermal energy and spin via the investigation of the relationship between thermal energy and the Pauli spin blockade.

In this paper, we investigate the current and the probabilities of the spin states of double quantum dots in series 
at different temperatures around the region where the Coulomb and Pauli spin blockades are observed.  
As the temperature  increases, 
it is demonstrated that the quadruplet Pauli spin blockade occurs as a result of the thermal energy 
and that the Coulomb and Pauli spin blockades coexist.   
Conversely, the triplet Pauli spin blockade is not clearly confirmed in the properties of the current flows.  
However, for the triplet Pauli spin blockade, 
the coexistence of the Coulomb and Pauli spin blockades is observed in the properties of the probability of the spin states of the double dots 
as a result of the thermally assisted Pauli spin blockade.
This indicates that the thermal energy enhances the current suppression effect 
caused by the high spin states that work as dark states, e.g., the triplet and quadruplet states.

\section{Current process for two and three levels}

\begin{figure}[h]
\includegraphics[width=1\columnwidth]{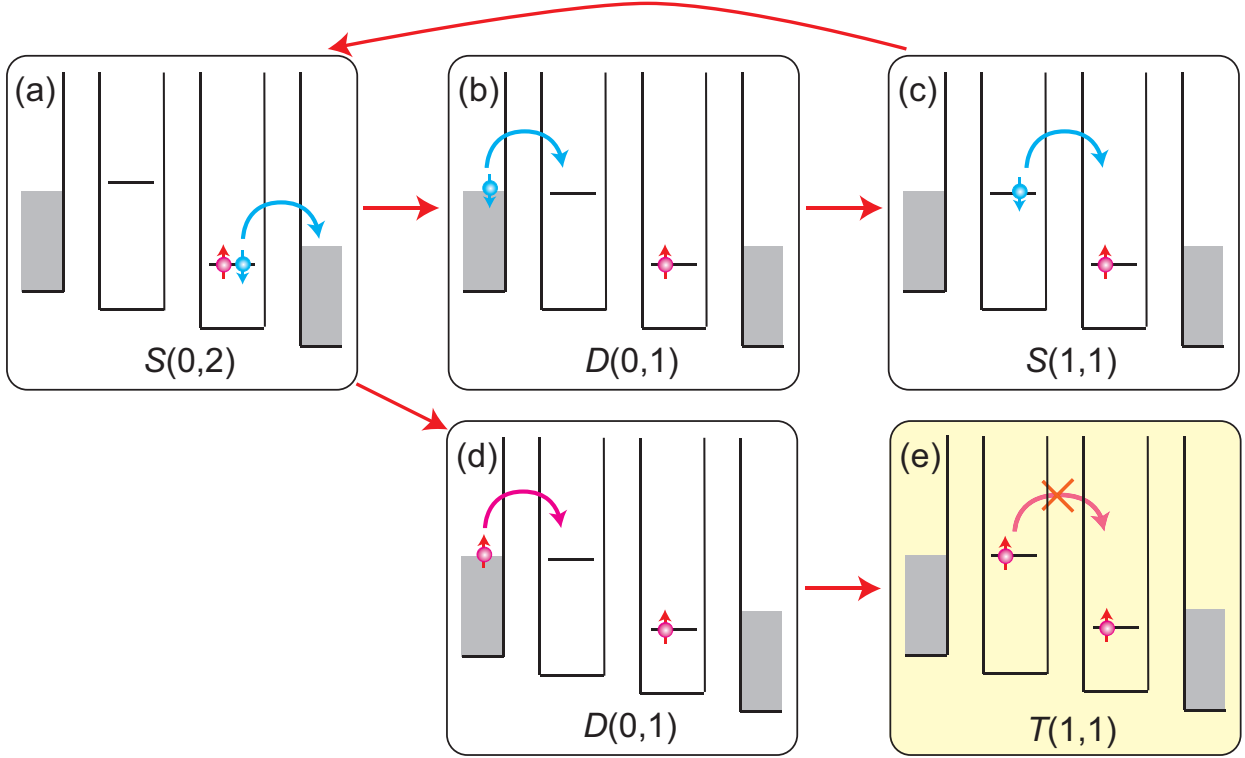}
\caption{
Schematic of the charge and spin states and their transitions in $(0, 2)\rightarrow(0, 1)\rightarrow(1, 1)\rightarrow(0, 2)$: 
(a) singlet state $S(0, 2)$, (b) doublet state $D(0, 1)$, 
(c) singlet state $S(1, 1)$, (d) doublet state $D(0, 1)$, and (e) triplet state $T(1, 1)$.  
Here, $(N_L, N_R)$ indicates the electron numbers in dots L and R, respectively. 
There are one and two levels in dots L and R, respectively. 
} 
\label{tpsb}
\end{figure}

First,  we discuss to the mechanisms of the triplet and quadruplet Pauli spin blockades. 
The charge transfer processes in $(0, 2)\rightarrow(0, 1)\rightarrow(1, 1)\rightarrow(0, 2)$, 
where the triplet Pauli spin blockade occurs,
is shown in Fig.~\ref{tpsb}. 
Here $(N_L, N_R)$ indicates the electron numbers in dots L and R, respectively.  
We start from the state in which two electrons are localized in dot R.  
Then, the ground state is the singlet state ($S=0$), $S(0, 2)$, 
where an electron with up  spin and another with down spin occupy the lowest single-particle level, as shown in Fig.~\ref{tpsb}(a).
$S$ indicates the total spin quantum number in the double quantum dots.

Suppose that the electron with down spin  tunnels into the drain electrode when a finite value of $V_{sd}$ is applied.    
Then, the doublet state $D(0, 1)$ is formed, as shown in Fig.~\ref{tpsb}(b). 
Here, if the electron with down spin tunnels into dot L from the source electrode, the singlet state $S(1, 1)$ 
is realized, as shown in Fig.~\ref{tpsb}(c). 
Finally, the electron with down spin tunnels into dot R 
and the double dot state returns to $S(0, 2)$. 
Therefore, in this case, the current flows through the double quantum dots.  
Note that this current cycle is correct, even if the roles of the up spin and down spin electrons are exchanged.
It is assumed that the electron with up spin in the source electrode tunnels into dot L after the formation of $D(0, 1)$, 
as shown in Fig.~\ref{tpsb}(d). 
Then, the triplet state $T(1, 1)$ is realized,  as shown in Fig.~\ref{tpsb}(e). 
In this situation, the electron with up spin in dot L is forbidden from tunneling into dot R according to the Pauli exclusion principle. 
Therefore, the current through the double quantum dots is suppressed. 
This indicates that the triplet state works as a dark state. 

\begin{figure}[h]
\includegraphics[width=1\columnwidth]{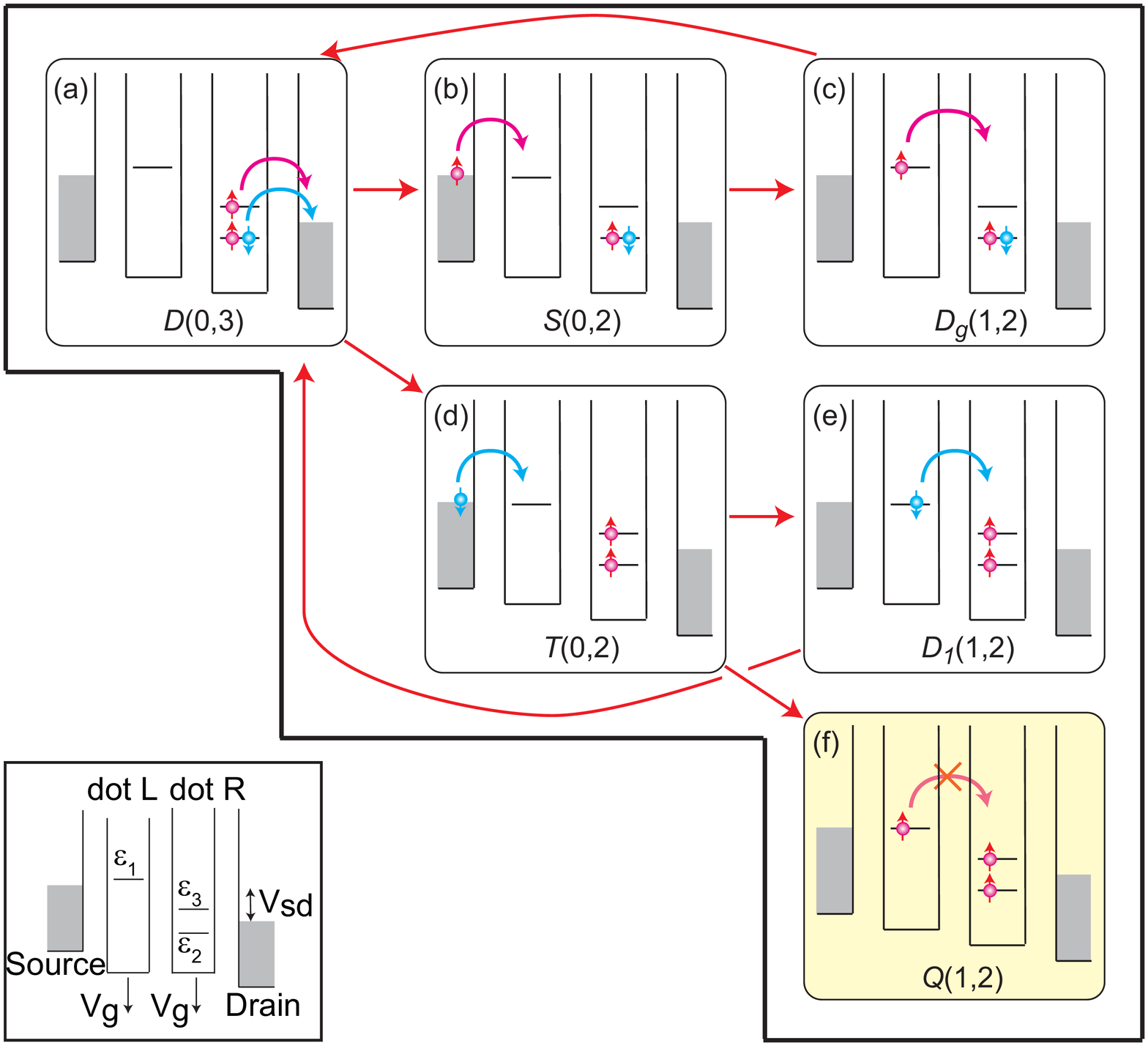}
\caption{
Schematic of the charge and spin states and their transitions in $(0, 3)\rightarrow(0, 2)\rightarrow(1, 2)\rightarrow(0, 3)$: 
(a) doublet state $D(0, 3)$, (b) singlet state $S(0, 2)$, 
(c) doublet state $D_g(1, 2)$, (d) triplet state $T(0, 2)$, (e) doublet state $D_1(1, 2)$, and (f) quadruplet state $Q(1, 2)$.  
Inset: Energy schematic of the double quantum dots in series used for the calculation.  
} 
\label{qpsb}
\end{figure}

Next, the charge transfer processes in $(0, 3)\rightarrow(0, 2)\rightarrow(1, 2)\rightarrow(0, 3)$, 
where the quadruplet Pauli spin blockade occurs,
is shown in Fig.~\ref{qpsb}. 
We start from the state in which three electrons are localized in dot R. 
Then, the ground state is the doublet state ($S=1/2$), $D(0,3)$,
 where two electrons (one with up  spin and the other with down spin) and 
one electron (e.g., with up spin) occupy the lowest  and the second lowest single-particle levels, respectively, 
as shown in Fig.~\ref{qpsb}(a).

When low $V_{sd}$ is applied to the double quantum dots, 
the electron with up spin, which occupies the second lowest single-particle level in dot R, 
tunnels into the drain electrode and the singlet state $S(0, 2)$ is formed, as shown in Fig.~\ref{qpsb}(b). 
Then, the electron with up spin tunnels into dot L from the source electrode and the doublet state $D_g(1, 2)$ is realized, 
as shown in Fig.~\ref{qpsb}(c). 
Finally, the electron with up spin tunnels into dot R 
and the double dot state returns to $D(0, 3)$. 
Therefore, in this case, the current flows through the double quantum dots.  

Next, let us consider the case in which a higher $V_{sd}$ is applied to the double quantum dots.  
Assuming that the electron with up spin, which occupies the lowest single-particle level in dot R, tunnels into the drain electrode, 
the electron with up spin, which occupies the second lowest single-particle level in dot R,  
transits to the lowest state and  the electron state in the double dots becomes $S(0,2)$, as shown  in Fig.~\ref{qpsb}(b), 
because of the short relaxation time of an electron with the same spin from the excited state to the ground state\cite{Huibers}. 
Then, the electron with up or down spin tunnels from the source electrode to dot L and the current flows through the double dots. 
Conversely, when  the electron with down spin, which occupies the lowest single-particle level in dot R, can tunnel into the drain electrode, 
the triplet state $T(0, 2)$ is realized as the first excited state, as shown in Fig.~\ref{qpsb}(d). 
Then, in the case in which the electron with down spin enters dot L, as shown in Fig.~\ref{qpsb}(e), 
the current flows through the double quantum dots. 
Conversely, when the electron with up spin enters dot L, 
the quadruplet state $Q(1, 2)$ is formed, as shown in Fig.~\ref{qpsb}(f).  
In this situation, the electron with up spin in dot L is forbidden from tunneling into dot R according to the Pauli exclusion principle; 
therefore, the current through the double quantum dots is suppressed, i.e., the quadruplet Pauli spin blockade occurs \cite{Amaha}. 
This indicates that the quadruplet state works as a dark state.

\section{Model and methods}

The calculation model of the double dots in series that we use is shown in the inset of Fig.~\ref{qpsb}. 
In the double quantum dots, dots L and R are coupled in series 
and the source and drain electrodes are connected to the two quantum dots. 
There is one level in dot L and two levels in dot R.

In this paper, we consider the following Hamiltonian: 
\begin{eqnarray}
H=H_{\rm DQD}+H_{\rm e}+H_{\rm e-d}, \nonumber
\end{eqnarray}

\begin{eqnarray}
H_{\rm DQD} &=&\sum_{i=1,\sigma=\uparrow, \downarrow}^{3}(\varepsilon_{i}-eV_{g})n_{i,\sigma}
+\sum_{i=1}^{3}U_in_{i,\uparrow}n_{i,\downarrow}\nonumber\\
&+&U_{12}n_1n_2+U_{23}n_2n_3+U_{31}n_3n_1\nonumber\\
&+&\sum_{\sigma=\uparrow, \downarrow}(t_{12}a^{\dag}_{1,\sigma}a_{2,\sigma}+t_{31}^{*}a^{\dag}_{1,\sigma}a_{3,\sigma}+{\rm H.c.}), \nonumber
\end{eqnarray}

\begin{eqnarray}
H_{\rm e} & =&  
  \sum_{k, \sigma=\uparrow, \downarrow} \varepsilon_{{\rm s}, k} n_{{\rm s} k,\sigma}
+ \sum_{k, \sigma=\uparrow, \downarrow} \varepsilon_{{\rm d}, k} n_{{\rm d} k,\sigma},\nonumber 
\end{eqnarray}
 
\begin{eqnarray}
H_{\rm e-d} & =  &
  \sum_{k, \sigma=\uparrow, \downarrow} v c^{\dag}_{{\rm s} k,\sigma} a_{1,\sigma}
+ \sum_{i=2}^3 \sum_{k, \sigma=\uparrow, \downarrow} v c^{\dag}_{{\rm d} k,\sigma} a_{i,\sigma}
+ {\rm H.c.},\nonumber
\end{eqnarray}
where $H_{\rm DQD}$ describes the electrons in the two quantum dots,
$H_{\rm e}$ describes the electrons in the source and drain electrodes,
and $H_{\rm e-d}$ describes the electron tunneling between the two electrodes and the dots.
Herein, we consider a single-particle level ($i = 1$) in dot L and two levels
(a lower level of $i = 2$ and a higher level of $i = 3$) in dot R. 
The energy states in the two dots are controlled by applying left and right gate voltages, $V_{gL}$ and $V_{gR}$, respectively.  Here, we assume that this model depicts vertical double quantum dots, 
because the quadruplet Pauli spin blockade has been observed only in vertical double quantum dots in series \cite{Amaha}. 
Therefore, we assume $V_{gL}=V_{gR}=V_{g}$ because $V_{gL}$ and $V_{gR}$ change the energy levels of dots L and R, respectively, 
nearly equally in vertical quantum double dots. 
Here $e$ is the elementary charge. 
Note that the following discussion holds true universally for double quantum dots, 
even though here we use parameter values that are appropriate for vertical double quantum dots. 
In addition, $\varepsilon_{i,\sigma}(i=1,2,3)$ are the single-particle energies of each
level, $U_i (i =1, 2, 3)$ are the intralevel Coulomb energies, $U_{ij} (i=1, 2, 3, i\neq j )$ are the interlevel Coulomb energies, 
and $t_{ij} (i, j = 1, 2, 3, i\neq j )$ are the tunnel coupling energies between the levels. 
$a_{i,\sigma}$, $a_{i,\sigma}^{\dag}$, and $n_{i,\sigma}$ ($i=1,2,3$, $\sigma=\uparrow, \downarrow$) are 
the annihilation, creation, and electron number operators of the three single-particle levels, respectively, 
and $\uparrow$, $\downarrow$ indicate the electron spins.  
$\varepsilon_{{\rm s}, k}$ ($\varepsilon_{{\rm d}, k}$), $c_{{\rm s} k,\sigma}$ ($c_{{\rm d} k,\sigma}$), 
$c^{\dag}_{{\rm s} k,\sigma}$ ($c^{\dag}_{{\rm d} k,\sigma}$), and $n_{{\rm s} k,\sigma}$ ($n_{{\rm d} k,\sigma}$)
are the energy,  annihilation, creation, and electron number operators  with the wavevector $k$ and the spin $\sigma$ in the source (drain) electrode, respectively. 
$v$ is the coupling energy between the single-particle levels in dot L (R) and the source (drain) electrode. 
We assume that the source drain voltage $V_{sd}$ symmetrically shifts the electro-chemical potentials 
in the source and drain electrodes,  $\mu_s=eV_{sd}/2$, and  $\mu_d=-eV_{sd}/2$, respectively. 

We diagonalized the Hamiltonian $H_{DQD}$ to solve for the spin states in the double dots, 
and calculated the current flow accounting for $H_{\rm e} $ and  $H_{\rm e-d}$ in the rate equations. 
The method to calculate the current flow is derived in Appendix A\cite{Fransson}.  

The parameters adopted for our calculation are based on a previous experiment \cite{Amaha}, 
and the following values were used:  
$U_1=U_2=U_3=3.5$ meV,  $U_{12}=2.1$ meV, $U_{23}=3.5$ meV, $U_{31}=1.75$ meV, 
$t_{12}=0.035$ meV, $t_{23}=0$ meV (levels 2 and 3 in the same dot), 
and $t_{31}=0.0035$ meV. 
The coupling strength between the dots and electrodes, $\gamma$, was 0.7 $\mu$eV (see Appendix A).

\section{Quadruplet Pauli spin blockade}
\label{sec:3level}

\begin{figure}[h]
\includegraphics[width=1\columnwidth]{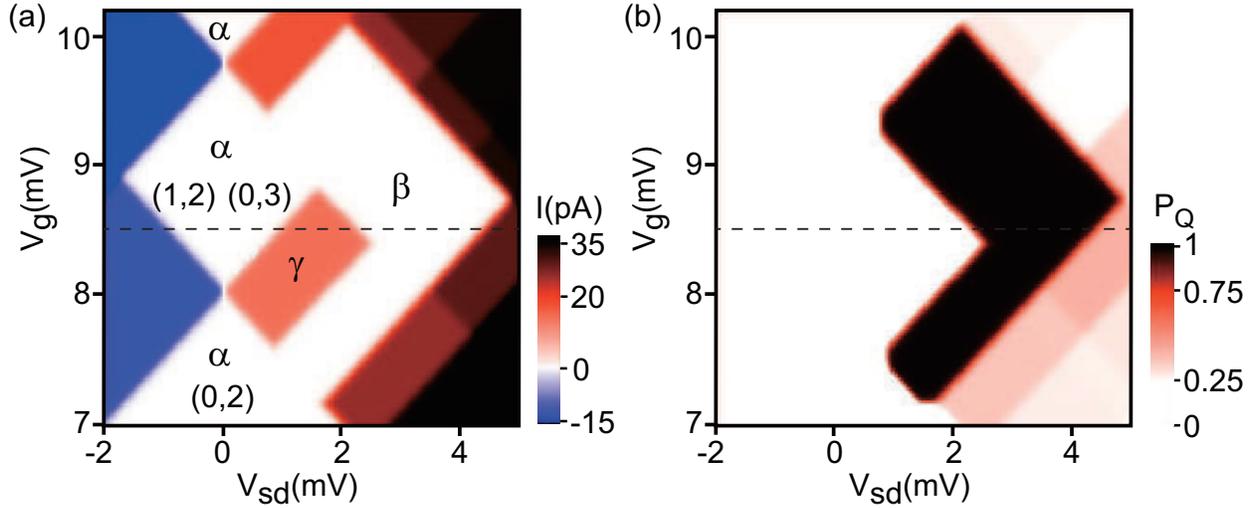}
\caption{
(a) Intensity plot of the current through the double quantum dots in series: $I$ as a function of 
the source drain voltage $V_{sd}$ and the gate voltage $V_{g}$ 
at a temperature of $T=$300 mK.  
 $(1,2)$ and $(0,3)$ are the charge configurations in the Coulomb diamond of the total electron number $N=3$, and $(0,2)$ is the Coulomb diamond of $N=2$. 
They are assigned according to the constant interaction model\cite{Amaha,Amaha2,Ota,Fuhrer,Amaha3}. 	
(b) Intensity plot of the probability of the quadruplet state, $P_{Q}$, 
as a function of $V_{sd}$ and $V_{g}$ at  $T=$300 mK. 
} 
\label{figure2}
\end{figure}

To study the influence of the thermal energy on the quadruplet Pauli spin blockade, 
we considered the three single-particle level model in the inset of Fig.~\ref{qpsb}. 
Here, we used the single-particle levels $\varepsilon_1-\varepsilon_2=3.85$ meV,  
and $\varepsilon_3-\varepsilon_2=0.3$ meV, based on $\varepsilon_2=0$ meV. 
Figure~\ref{figure2}(a) shows an intensity plot of the current through the double quantum dots in series $I$ 
as a function of the source drain voltage $V_{sd}$ and  the gate voltage  $V_{g}$ at a temperature of $T=$300 mK. 
Similar to the two-level case, the Coulomb staircase is clearly observed 
because of the sufficiently low temperature 
and  the current flow is suppressed in several regions in Fig.~\ref{figure2}(a). 
The current-suppressed rhombuses ($\alpha$ in panel (a)) with centers  focused on $V_{sd}=0$ V correspond to the Coulomb blockade regions.

Conversely, the current-suppressed region for positive values of $V_{sd}$, shown as the $\beta$ region in Fig.~\ref{figure2}(a), 
is the quadruplet Pauli spin blockade region \cite{Amaha}. 
Therefore, the quadruplet state works as the dark state. 
In the region of the rectangular conductive island marked as the $\gamma$ region  in Fig.~\ref{figure2}(a), 
the current flows via  $S(0,2)$, $T(0,2)$, and $D_g(1,2)$, $D_1(1,2)$, or $D(0,3)$, as indicated in Fig.~\ref{qpsb}.

Here, we show an intensity plot of the probability of the quadruplet state of the electron state in the double dots being Q(1,2), 
$P_{Q}$, as a function of $V_{sd}$ and $V_{g}$ at $T=$300 mK in Fig.~\ref{figure2}(b). 
The value of  $P_{Q}$ is nearly zero at negative values of  $V_{sd}$. 
Accordingly, at negative values of  $V_{sd}$, the quadruplet Pauli spin blockade does not occur. 
Similarly, in the regions corresponding to $\alpha$ and $\gamma$ in Fig.~\ref{figure2}(a), 
the value of $P_{Q}$ is nearly zero because the energy necessary to form the quadruplet state is not available. 
Conversely, the  value of  $P_{Q}$ in the region corresponding to $\beta$ in Fig.~\ref{figure2}(a) is nearly unity; 
therefore, the quadruplet Pauli spin blockade occurs. 
This shows that the high spin state $Q(1, 2)$ is achieved via the dead-end path  (a)$\rightarrow$(d)$\rightarrow$(f) in Fig.~\ref{qpsb} 
and that the current is suppressed as a result of the realization of $Q(1, 2)$. 
When $V_{sd}$ increases further, 
the value of  $P_{Q}$ is not unity but is a finite value. 
In this region, the current can flow via the quadruplet state 
because two electrons can contribute to the current flow simultaneously \cite{Hatano2}. 

\begin{figure}[h]
\includegraphics[width=1\columnwidth]{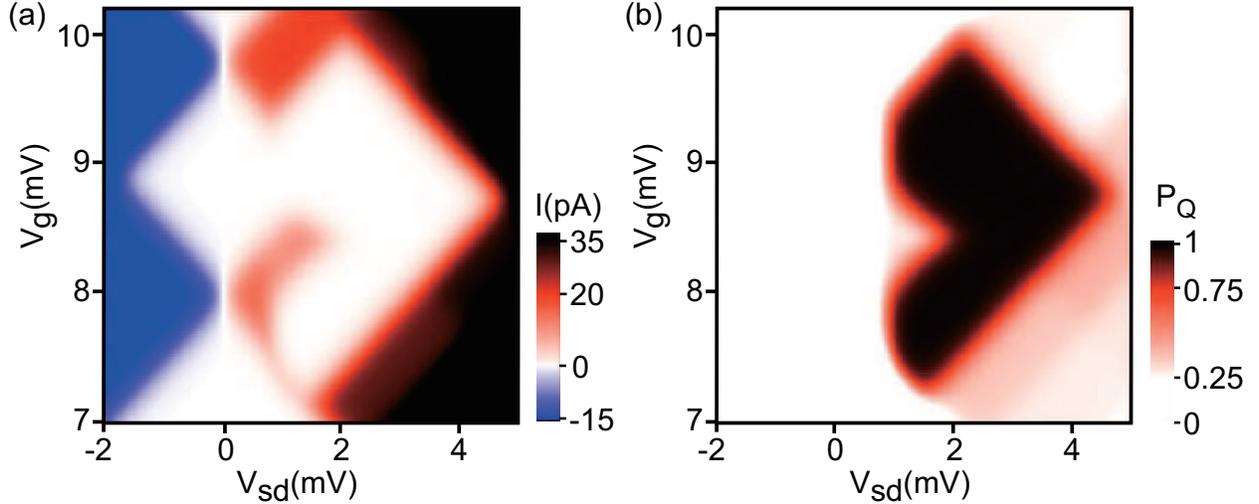}
\caption{
(a) Intensity plot of $I$ as a function of $V_{sd}$ and $V_{g}$ 
at $T=1$ K.  (b) Intensity plot of  $P_{Q}$ as a function of $V_{sd}$ and $V_{g}$ at  $T=1$ K. 
} 
\label{fig1k}
\end{figure}

 In Fig.~\ref{fig1k}(a), we show an intensity plot of $I$ as a function of $V_{sd}$ and $V_{g}$ at $T=1$ K. 
Compared to Fig.~\ref{figure2}(a), the value of $I$ changes somewhat gradually with $V_{sd}$ and $V_{g}$ 
as a result of the increase in the thermal energy. 
However, the  $\gamma$ region becomes small and the $\beta$ region, i.e., quadruplet Pauli spin blockade region, 
becomes larger than that in Fig.~\ref{figure2}(a).   
We show an intensity plot of  $P_{Q}$ as a function of $V_{sd}$ and $V_{g}$ at $T=1$ K in Fig.~\ref{fig1k}(b). 
As $T$ increases, the edges of the region of  $P_{Q}\sim 1$ are smeared 
as a result of the thermal energy compared to that at $T=300$ mK. 
However, the region of $P_{Q}\sim 1$  extends to the regions of 
the Coulomb blockade ($\alpha$) and the rectangular conductive island ($\gamma$) in Fig.~\ref{figure2}(a). 

\begin{figure}[h]
\includegraphics[width=1\columnwidth]{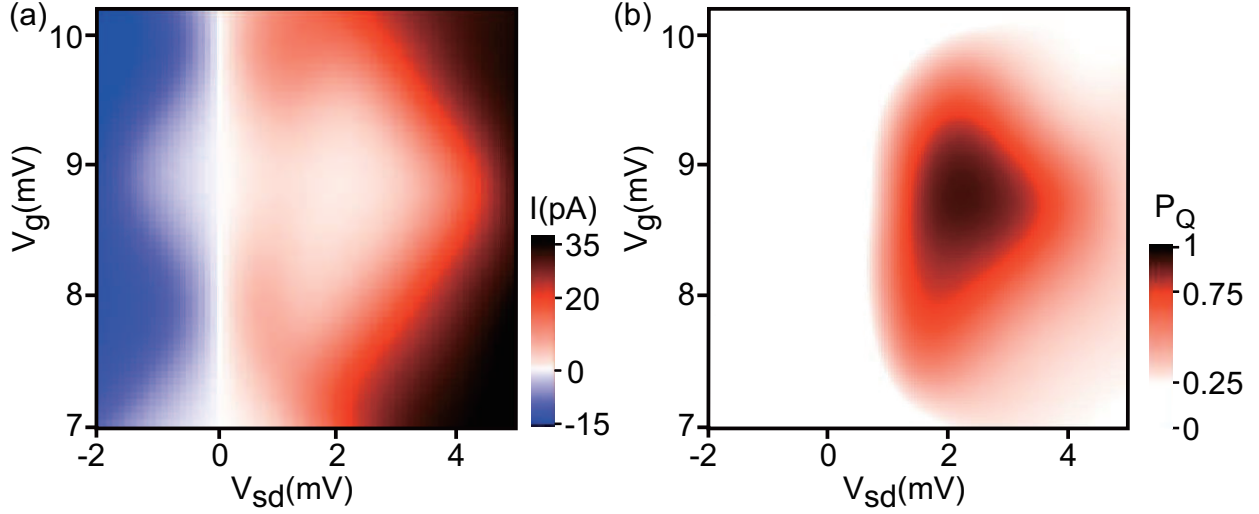}
\caption{
Intensity plot of $I$ as a function of $V_{sd}$ and $V_{g}$ 
at $T=$3 K.  (b) Intensity plot of  $P_{Q}$ as a function of $V_{sd}$ and $V_{g}$ at  $T=$3 K. 
} 
\label{figure3}
\end{figure}

Next, we show an intensity plot of $I$ as a function of $V_{sd}$ and $V_{g}$ at $T=$3 K in Fig.~\ref{figure3}(a). 
Compared to Fig.~\ref{fig1k}(a), the value of $I$ changes more gradually with $V_{sd}$ and $V_{g}$ as a result of the increase in
the thermal energy and the Coulomb blockade region is observed. However, the rectangular conductive
island  ($\gamma$) in Fig.~\ref{figure2}(a) cannot be definitively confirmed. In addition, we show an intensity plot of $P_{Q}$ as a
function of $V_{sd}$ and $V_{g}$ at $T=$3 K in Fig.~\ref{figure3}(b). 
As $T$ increases, the region of $P_{Q}\sim 1$ is even more smeared than
at $T=1$ K as a result of the thermal energy and becomes triangular in shape. Therefore, the Coulomb
blockade ($\alpha$), the quadruplet Pauli spin blockade ($\beta$), 
and the rectangular conductive island ($\gamma$) in Fig.~\ref{figure2}(a) cannot be distinguished. 
This indicates that the quadruplet Pauli spin blockade occurs in the region where it
does not occur at $T=300$ mK, i.e., in the $\alpha$ and $\beta$ regions in Fig.~\ref{figure2}(a). 
Moreover, the Pauli spin and Coulomb blockades coexist in part of the $\alpha$ region in Fig.~\ref{figure2} (a) 
as a result of the thermal energy.

\begin{figure}[h]
\includegraphics[width=1\columnwidth]{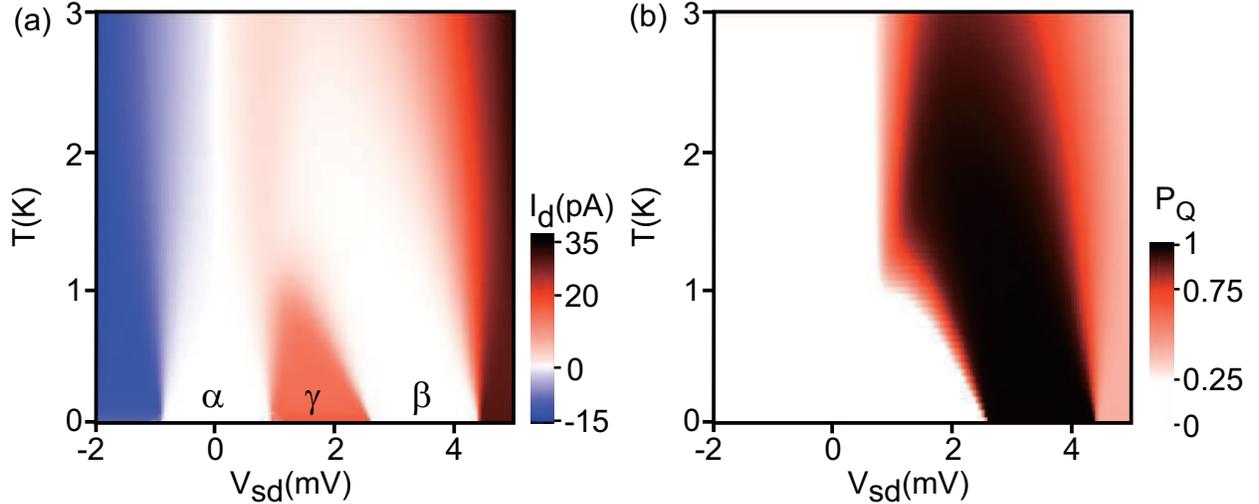}
\caption{
(a) Intensity plot of $I$ as a function of $T$ and $V_{sd}$ 
at $V_{g}=8.5$ meV.  
(b) Intensity plot of $P_{Q}$ as a function of $T$ and $V_{sd}$ at $V_{g}=8.5$ meV.  
} 
\label{figure4}
\end{figure}

In Fig.~\ref{figure4}(a), we show an intensity plot of $I$ as a function of $T$ and $V_{sd}$ at fixed $V_{g}=$8.5 mV, 
to investigate the temperature dependence of the current suppression of $I$ in the $\gamma$ region  
in Fig.~\ref{figure2}(a). 
The $V_{g}$ sweeping direction is indicated 
by the dashed line in Fig.~\ref{figure2}(a), and the $\alpha$, $\beta$, and $\gamma$ regions correspond to those in Fig.~\ref{figure2}(a). 
As the temperature increases from 0 K to 1 K, the $\gamma$ region abruptly diminishes and 
the quadruplet Pauli spin blockade region, $(\beta)$, abruptly extends in the direction of the lower values of $V_{sd}$. 
Figure~\ref{figure4}(b) shows an intensity plot of  $P_{Q}$  as a function of $T$ and $V_{sd}$ at fixed $V_{g}=$8.5 mV.  
Similar to Fig.~\ref{figure4}(a), the region of $P_{Q}\sim 1$ extends in the direction of lower values of $V_{sd}$, 
as the temperature increases from 0 K to 1 K.
Therefore, the quadruplet Pauli spin blockade occurs for lower values of $V_{sd}$ with increasing temperature.  
This indicates that the thermally assisted quadruplet Pauli spin blockade occurs. 
For normal semiconductors, as the temperature increases, 
the quantum mechanical effects become weaker and the properties caused by these effects are smeared. 
This current suppression with increasing temperature is different from the normal semiconductor-temperature dependence.

\begin{figure}[h]
\includegraphics[width=1\columnwidth]{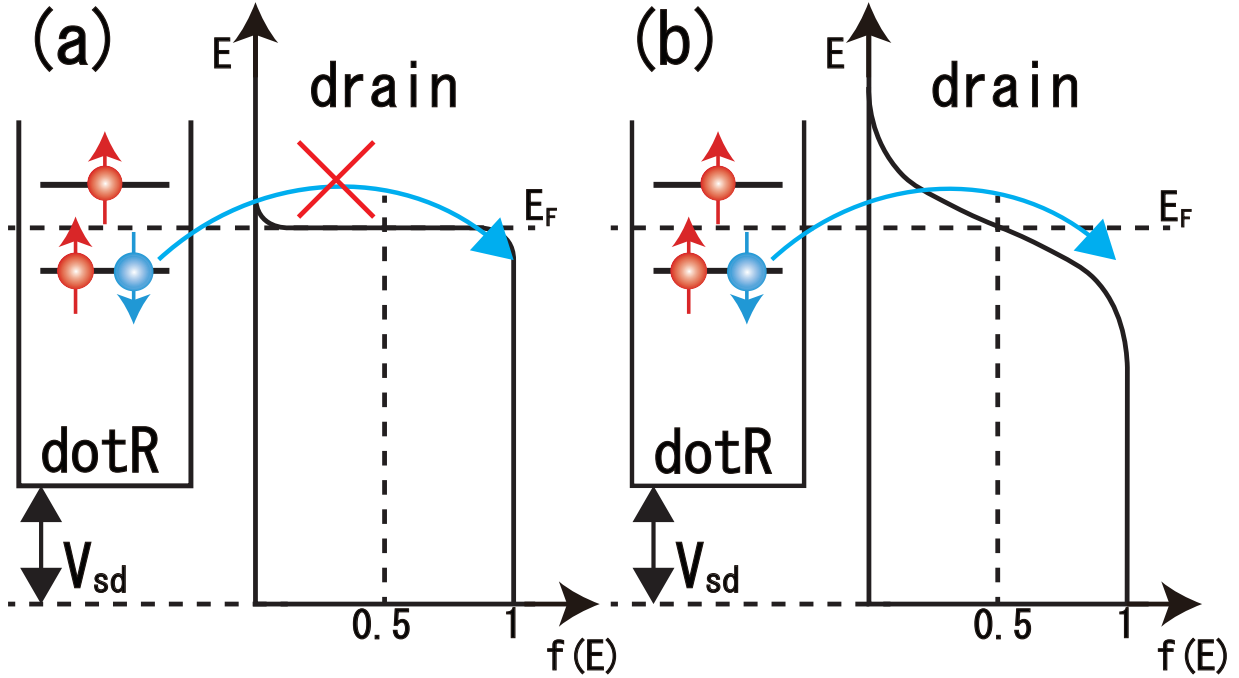}
\caption{
Energy schematic of the dot R and drain at  (a) low and (b) high temperatures.  
} 
\label{figure5}
\end{figure}

To reveal the mechanism of the thermally assisted quadruplet Pauli spin blockade,
we show schematic energy diagrams of dot R and the drain electrode at low and high temperatures 
in Figs.~\ref{figure5}(a) and \ref{figure5}(b), respectively. 
Here, the three-electron state of dot R is $D(0, 3)$, as shown in Fig.~\ref{qpsb}(a).  
To generate the quadruplet Pauli spin blockade, 
it is necessary that the electron with the down spin tunnels from dot R to the drain electrode
and that $T(0, 2)$ is realized as shown in Fig.~\ref{qpsb}(d). 
For low temperatures, 
the Fermi distribution function of the drain electrode abruptly changes near the Fermi energy, 
as indicated in Fig.~\ref{figure5}(a). 
Accordingly, there are hardly any empty states just below the Fermi energy 
and the electron with down spin in dot R is forbidden from tunneling into the drain electrode. 
A high value of $V_{sd}$ is needed for the electron with down spin 
to tunnel from dot R to the drain electrode and to realize $T(0, 2)$; 
therefore, a high value of $V_{sd}$ needs to be applied to the double quantum dots to cause the quadruplet Pauli spin blockade.

However, as the temperature increases, 
the Fermi distribution function of the drain gradually changes near the Fermi energy 
and a large number of empty states exist just below this energy, as indicated in Fig.~\ref{figure5}(b). 
As a consequence, it is possible for an electron with down spin to tunnel from dot R to the drain electrode 
and the quadruplet state is realized even for comparatively lower values of $V_{sd}$. 
Therefore, assisted by the thermal energy, the quadruplet Pauli spin blockade occurs.

Here, we discuss the difference between the triplet and quadruplet Pauli spin blockades. 
For the quadruplet Pauli spin blockade, because of the current island located between the Coulomb and  quadruplet Pauli spin blockades, it is easy to observe the thermally assisted quadruplet Pauli spin blockade. 
Conversely, for the triplet Pauli spin blockade, the region of this blockade is adjacent to that of the Coulomb blockade.  
Therefore, we cannot clearly observe the thermally assisted triplet Pauli spin blockade.  
In both Pauli spin blockade cases, 
one can confirm the coexistence of the quadruplet Pauli spin and Coulomb blockades 
via a measurement of not only the current flow but also the spin states in the double quantum dots.  

\section{Triplet Pauli spin blockade}
\label{sec:2level}

\begin{figure}
\includegraphics[width=1\columnwidth]{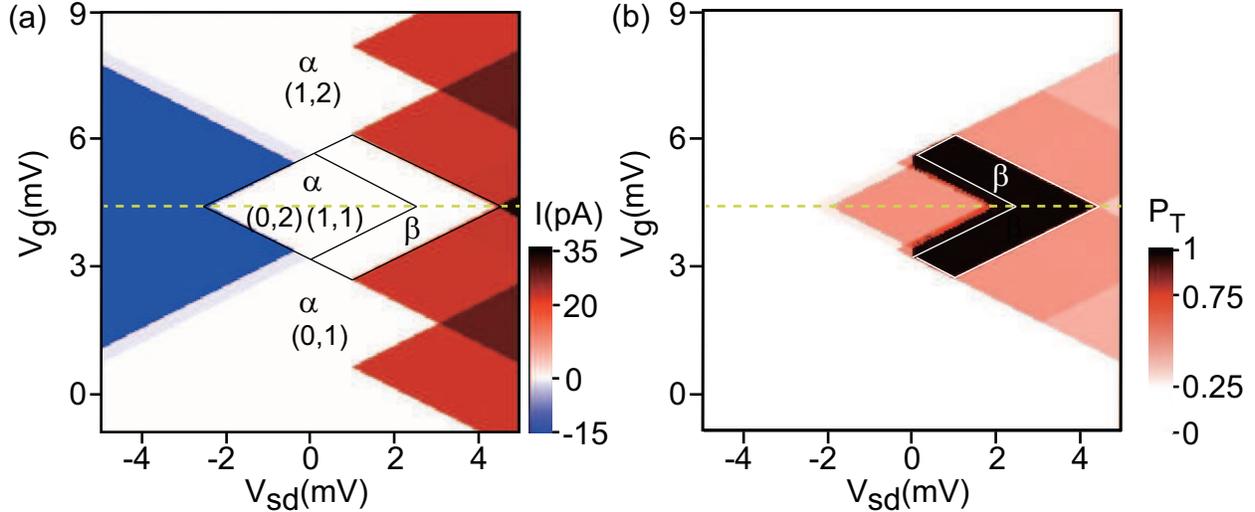}
\caption{
(a) Intensity plot of $I$ 
as a function of $V_{sd}$ and $V_{g}$ 
at $T=100$ mK.  
$(1,1)$ and $(0,2)$ are the charge configurations in the Coulomb diamond of the total electron number $N=2$, and $(0,1)$ is the Coulomb diamond of $N=1$. 
They are assigned according to the constant interaction model\cite{Amaha,Amaha2,Ota,Fuhrer,Amaha3}. 
(b) Intensity plot of  the probability of the triplet state $T(1,1)$, $P_{T}$, 
as a function of $V_{sd}$ and $V_{g}$ at  $T=100$ mK. 
} 
\label{cur2level100mk}
\end{figure}

How does the thermal energy affect the triplet Pauli spin blockade?
We consider a model, in which there is one single-particle level in each dot. 
Here, we assume  $\varepsilon_1-\varepsilon_2=1.05$ meV and $\varepsilon_3-\varepsilon_2=350$ meV
based on $\epsilon_2=0$ meV for the calculation model in the inset of  Fig.~\ref{qpsb}. 
Then, the energy level, $\epsilon_3$ can be excluded because it is much larger than the applied $V_{sd}$ and $V_g$. 
Figure~\ref{cur2level100mk}(a) shows an intensity plot of $I$ 
as a function of $V_{sd}$ and $V_{g}$ at $T=$100 mK. 
The Coulomb staircase is clearly observed 
because of the sufficiently low temperature, and
the current flow is suppressed in several regions. 
The Coulomb blockade ($\alpha$) and triplet Pauli spin blockade ($\beta$) regions are confirmed in Fig.~\ref{cur2level100mk}(a) \cite{Ono}.   
Note that the property of $I$ is symmetric with respect to $V_{g}=4.35$ mV 
as a result of the electron hole symmetry. 

To investigate the electron states of the double dots in the region where the current is suppressed, 
we show an intensity plot of the probability of the triplet state of the electron state in the double dots being $T(1,1)$, $P_{T}$, 
as a function of  $V_{sd}$ and $V_{g}$ at a temperature of $T=$100 mK in Fig.~\ref{cur2level100mk}(b). 
The $\beta$ region, where the triplet Pauli spin blockade occurs,  is surrounded by a white line 
and corresponds to the triplet Pauli spin blockade region in  Fig.~\ref{cur2level100mk}(a).
The value of $P_{T}$ is nearly unity for the $\beta$ region.  
This shows that the high spin state $T(1, 1)$ is achieved via the dead-end path  (a)$\rightarrow$(d)$\rightarrow$(e) in Fig.~\ref{tpsb} 
and that the current is suppressed as a result of the realization of $T(1, 1)$. 
For the $\alpha$ region, where the Coulomb blockade occurs, 
$P_{T}$ has a nonzero finite value,even though $T(1, 1)$ is the excited state \cite{memo1}. 
This is because a finite value of $V_{sd}$ is applied, i.e., this system is in non-equilibrium. 
Further, the region of $P_{T}\sim 1$ extends into the edge of the $\alpha$ region at $V_{sd}>0$ in Fig.~\ref{cur2level100mk}(a); 
therefore, in the region of $P_{T}\sim 1$ in the Coulomb blockade region, 
the triplet Pauli spin blockade coincides with the Coulomb blockade. 

\begin{figure}
\includegraphics[width=1\columnwidth]{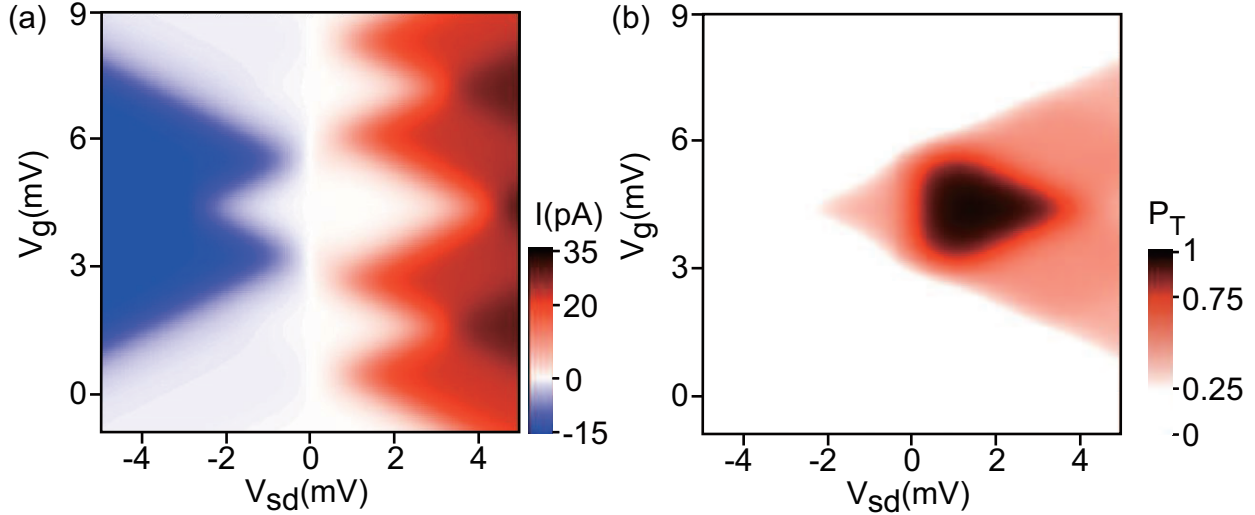}
\caption{
(a) Intensity plot of $I$ as a function of $V_{sd}$ and $V_{g}$ 
at $T=$3 K.  (b) Intensity plot of  $P_{T}$ 
as a function of $V_{sd}$ and $V_{g}$ at  $T=$3 K. 
} 
\label{cur2level3k}
\end{figure}

In Fig.~\ref{cur2level3k}(a), we show  an intensity plot of $I$ 
as a function of $V_{sd}$ and $V_{g}$ at $T=$3 K. 
Compared to Fig.~\ref{cur2level100mk}(a), 
the value of $I$ gradually changes with $V_{sd}$ and $V_{g}$ as a result of the increase in the thermal energy.  
The current suppression regions, the Coulomb and triplet Pauli spin blockades, are observed. 
Here, we show an intensity plot of  $P_{T}$ 
 as a function of $V_{sd}$ and $V_{g}$ at $T=$3 K in Fig.~\ref{cur2level100mk}(b). 
As $T$ increases, the edges of the region of  $P_{T}\sim 1$ are smeared 
as a result of thermal energy compared to those at $T=100$ mK. 
However, $P_{T}$ is nearly unity for $V_{sd}>0$ mV in the $\alpha$ region in Fig.~\ref{cur2level100mk}(a), 
as opposed to $P_{T}$ in Fig.~\ref{cur2level100mk}(b).  
This shows that the high spin state $T(1, 1)$ is achieved in the region for $V_{sd}>0$ mV in the Coulomb blockade region 
as the temperature increases. 
This indicates that the dead-end path  (a)$\rightarrow$(d)$\rightarrow$(e) in Fig.~\ref{tpsb} 
is caused  by the  thermal energy and that the Coulomb and  Pauli spin blockades coexist.  
Therefore, in the region of $P_{T}\sim 1$ in the Coulomb blockade region, the simple Coulomb blockade does not occur; 
however, an electron is forbidden from tunneling through dot R 
as a result of both the Coulomb repulsion and the Pauli exclusion principle in the double dots. 
However, this cannot be observed from the properties of $I$; 
to confirm this, the spin state in the double quantum dots needs to be detected.

\section{Conclusions}

We calculated the current and spin states in double quantum dots in series. 
As the temperature  increases, 
the quadruplet Pauli spin blockade occurs as a result of the thermal energy. 
The triplet Pauli spin blockade is not clearly confirmed 
in the properties of the current flows.  
However, for both Pauli spin blockades, 
the coexistence of the Coulomb and Pauli spin blockades occurs as a result of the thermally assisted Pauli spin blockade
and is reflected in the properties of the probability of the spin states of the double dots.
This coexistence affects the initialization and readout of qubits in quantum dot quantum computing 
and may reduce the qubit fidelity.  
Moreover, in nanoscale devices, heat transport and dissipation have become important topics \cite{Pop}, 
e.g., the operation temperatures of quantum devices and quantum computing \cite{Ono2} and a single-quantum-dot heat valve \cite{Dutta}.  
Therefore,  
our results pave a way toward the clarification of spin based transport and thermal excitation in nanoscale devices, 
and it may also be possible to realize novel quantum devices, e.g., quantum heat engine \cite{Ono3}, using the thermally assisted Pauli spin blockade.

\begin{acknowledgments}
We thank M. Ikeda, S. Kawamagari, D. Annaka, H. Itoh, T. Kato, and A. Kowata for discussions and comments. 
T.H. acknowledges the Research Grant of the College of Engineering, Nihon University. 
W. I. acknowledges support from KAKENHI Grant Nos. JP15K05118 and JP18H04282.
\end{acknowledgments}

\appendix
\section{Details of calculations}

\begin{table}[h]
\caption{Basis vectors to diagonalize $H_{DQD}$. 
$N(=N_L+N_R)$ and $S$ indicate the total electron numbers 
and total spin quantum number in the double quantum dots, respectively. 
$|\xi_1,\xi_2, \xi_3\rangle$ represents the occupation of the single-particle energy levels, 
and $\xi_i$ denotes the local states, 0, $\uparrow$, $\downarrow$, and $\uparrow\downarrow$, which stand 
for empty, spin states up and down of the occupied electrons, and two electron occupation. 
$\sigma$ indicates $\uparrow$ or $\downarrow$, and 
$\bar{\sigma}$ represents $\downarrow (\uparrow)$ for $\sigma=\uparrow (\downarrow)$
}
\label{table1}
\centering
\begin{tabular}[c]{|c|c|c|c|c|}\hline
$N$ & $S$ & Basis vectors \\ \hline
0 & 0 & $|0, 0, 0\rangle$ \\ \hline
1 & $\frac{1}{2}$ & $|\sigma, 0, 0\rangle$, $|0, \sigma, 0 \rangle$, $|0, 0, \sigma \rangle$\\ \hline
2 & 0 & $|\uparrow\downarrow, 0, 0\rangle$, $|0, \uparrow\downarrow, 0\rangle$, 
$|0, 0, \uparrow\downarrow\rangle$, $\frac{1}{\sqrt{2}}(|\uparrow, \downarrow, 0\rangle-|\downarrow, \uparrow, 0\rangle)$\\ 
& & 
$\frac{1}{\sqrt{2}}(|0, \uparrow, \downarrow\rangle-|0, \downarrow, \uparrow\rangle)$, 
$\frac{1}{\sqrt{2}}(|\uparrow, 0, \downarrow\rangle-|\downarrow, 0, \uparrow\rangle)$\\ \hline
2 & 1 & $|\sigma, \sigma, 0\rangle$, $|0, \sigma, \sigma\rangle$, $|\sigma, 0, \sigma\rangle$, 
$\frac{1}{\sqrt{2}}(|\uparrow, \downarrow, 0\rangle+|\downarrow, \uparrow, 0\rangle)$\\ 
& & 
$\frac{1}{\sqrt{2}}(|0, \uparrow, \downarrow\rangle+|0, \downarrow, \uparrow\rangle)$, 
$\frac{1}{\sqrt{2}}(|\uparrow, 0, \downarrow\rangle+|\downarrow, 0, \uparrow\rangle)$\\ \hline
3 & $\frac{1}{2}$ & $|\sigma, 0, \uparrow\downarrow\rangle$, $|0, \sigma, \uparrow\downarrow \rangle$, 
$|0, \uparrow\downarrow, \sigma\rangle$ \\
& & $|\sigma, \uparrow\downarrow,0\rangle$, $|\uparrow\downarrow, \sigma, 0 \rangle$, 
$|\uparrow\downarrow, 0, \sigma\rangle$ \\
& & 
$\frac{1}{\sqrt{2}}(|\sigma, \uparrow, \downarrow \rangle -|\sigma, \downarrow, \uparrow\rangle$) \\
& & 
$\frac{1}{\sqrt{6}}(|\sigma, \uparrow, \downarrow \rangle +|\sigma, \downarrow, \uparrow\rangle
- 2|\bar{\sigma}, \sigma, \sigma\rangle)$\\ 
\hline
3 & $\frac{3}{2}$ & $|\sigma, \sigma, \sigma\rangle$, 
$\frac{1}{\sqrt{3}}(|\uparrow, \uparrow, \downarrow\rangle+|\uparrow, \downarrow, \uparrow\rangle
+|\downarrow, \uparrow, \uparrow\rangle)$ \\
& &
$\frac{1}{\sqrt{3}}(|\downarrow, \downarrow, \uparrow\rangle+|\downarrow, \uparrow, \downarrow\rangle
+|\uparrow, \downarrow, \downarrow\rangle)$\\ \hline
4 & 0 & $|\uparrow\downarrow, \uparrow\downarrow, 0 \rangle$,
$|\uparrow\downarrow, 0, \uparrow\downarrow \rangle$, 
$|0, \uparrow\downarrow, \uparrow\downarrow \rangle$ \\
& & 
$\frac{1}{\sqrt{2}}(|\uparrow\downarrow, \uparrow, \downarrow\rangle-|\uparrow\downarrow, \downarrow, \uparrow\rangle)$
\\ 
& & 
$\frac{1}{\sqrt{2}}(|\uparrow, \uparrow\downarrow, \downarrow\rangle-|\downarrow, \uparrow\downarrow, \uparrow\rangle)$
\\ 
& & 
$\frac{1}{\sqrt{2}}(|\uparrow, \downarrow, \uparrow\downarrow\rangle-|\downarrow, \uparrow, \uparrow\downarrow\rangle)$
\\ \hline
4 & 1 & $|\uparrow\downarrow, \sigma, \sigma\rangle$, $|\sigma, \uparrow\downarrow, \sigma\rangle$, 
$|\sigma, \sigma, \uparrow\downarrow\rangle$\\
& & $\frac{1}{\sqrt{2}}(|\uparrow\downarrow, \uparrow, \downarrow\rangle+|\uparrow\downarrow, \downarrow, \uparrow\rangle)$
\\
& &
$\frac{1}{\sqrt{2}}(|\uparrow, \uparrow\downarrow, \downarrow\rangle+|\downarrow, \uparrow\downarrow, \uparrow\rangle)$
\\
& & 
$\frac{1}{\sqrt{2}}(|\uparrow, \downarrow, \uparrow\downarrow\rangle+|\downarrow, \uparrow, \uparrow\downarrow\rangle)$
\\ \hline
5 & $\frac{1}{2}$ & $|\sigma, \uparrow\downarrow, \uparrow\downarrow\rangle$, 
$|\uparrow\downarrow, \sigma, \uparrow\downarrow\rangle$, $|\uparrow\downarrow, \uparrow\downarrow, \sigma\rangle$ 
\\ \hline
6 & 0 & $|\uparrow\downarrow, \uparrow\downarrow, \uparrow\downarrow\rangle$
\\ \hline
\end{tabular}
\end{table}

We describe the calculation method of the probability of the quadruplet and triplet states, $P_Q$ and $P_T$, 
respectively, and the current flow, $I$.  
It is assumed that the coupling between two quantum dots is relatively strong, 
and that the coupling between the double quantum dot and two electrodes 
is a sequential tunneling process\cite{Bruus}. 
Therefore, we diagonalize the Hamilton $H_{DQD}$
and obtain the eigenvalues and eigenvectors of $H_{DQD}$. 
The eigenvectors can be indicated by $N(=N_L+N_R)$ and $S$, which  are the total electron 
and total spin numbers in the double quantum dot respectively, because $N$ and $S$ are 
conserved quantities for $H_{DQD}$. 
In Table \ref{table1}, we show the basis vectors to use for the digitalization of $H_{DQD}$, 
corresponding to the eigenvectors of  $H_{DQD}$ at $t_{1i}=t^*_{i1}=0, (i=2,3)$. 
Therefore, the eigenvectors of $H_{DQD}$ are expressed by a linear combination of the basis vectors. 

Thereafter, we calculate the transition probability from the initial spin state, $|i\rangle$, to the final spin state, 
$|f\rangle$, in the double quantum dot, $\Gamma_{fi}^{s(d)}$,
corresponding to the coupling strength between the double quantum dot and source (drain) electrode, as follows: 
\begin{eqnarray}
\Gamma_{fi}^{s(d)}=\frac{2\pi}{\hbar}\sum_{I,F}|\langle F |\langle f|H_{e-d}|i\rangle |I\rangle|^2W_I
\delta(E_{fF}-E_{iI}).
\end{eqnarray}
Here  $|I \rangle$ and $|F \rangle$ are the initial and final states in the source (drain) electrode, 
respectively. 
$|i \rangle$ and $|f \rangle$ are  the eigenvectors of $H_{DQD}$ and are denoted using $N$ and $S$.  
 $W_I$ is the probability that the initial state in the source (drain) electrode is $|I \rangle$ under thermal equilibrium. 
$E_{iI}$ and $E_{fF}$ are the total energies of  the initial and final states  of the double quantum dot and two electrodes, respectively. 
$\Gamma_{fi}^{s(d)}$ is nonzero for $N_f=N_i\pm 1$ and is numerically obtained in our calculation. 
$N_i$ and $N_f$ are the numbers of electrons in the double quantum dot for $|i\rangle$ and $|f\rangle$, 
 respectively. 

As an example, we show $\Gamma_{21}^{s}$, 
corresponding to the transition probability from the initial spin state $|1\rangle$ to 
the final state $|2\rangle$. 
$|1\rangle$ is the state in which there are no electrons in the double quantum dot, 
and $|2\rangle$ is the state in which there is one electron  in the double quantum dot, 
as follows:  
\begin{eqnarray}
|1\rangle&=&|0,0,0\rangle\nonumber\\
|2\rangle&=&A|\uparrow,0,0\rangle+B|0,\uparrow,0\rangle+C|0,0,\uparrow\rangle, \nonumber
\end{eqnarray}
where $A$, $B$ and $C$ are coefficients ($|A|^2+|B|^2+|C|^2=1$). 
Therefore, $\Gamma_{21}^{s}$ is calculated as follows:  
\begin{eqnarray}
\Gamma_{21}^{s}&=&\frac{2\pi}{\hbar}|\langle 2|H_{e-d}|1\rangle|^2W_I\delta(E_{fF}-E_{iI})\nonumber\\
&=&\frac{2\gamma}{\hbar}f_{s}(E_{2}-E_{1})|A|^2,
\end{eqnarray}
where $\gamma=\pi v^2\sum_{k}\delta(E_{2}-E_{1}-\varepsilon_{sk})(=\pi v^2\sum_{k}\delta(E_{2}-E_{1}-\varepsilon_{dk}))$ is 
the tunneling strength between the source  electrode and the double quantum dot. 
$f_{s}(E_{2}-E_{1})$ is the Fermi distribution function in the source electrode,
and $E_{2}$ and $E_{1}$ are the eigenvalues of $|2\rangle$ and $|1\rangle$, respectively. 
$\varepsilon_{s(d)k}$ is the energy of the source  (drain) electrode. 
Similarly, we define $f_d(\varepsilon_{dk})$  
as the Fermi distribution function in the drain electrode. 

In addition, we define the transition probability from $|i\rangle$ to $|f\rangle$, $\Gamma_{fi}$, 
which is contributed to by tunneling between the source electrode to the double quantum dot
and by tunneling between the drain electrode and the double quantum dot,
 as follows:  
\begin{eqnarray}
\Gamma_{fi}=\Gamma_{fi}^{s}+\Gamma_{fi}^{d}.
\end{eqnarray}

We denote the probability of the eigenvector $|i\rangle$ as $P(i)$, and $\Gamma_{fi}$, 
the rate equation is expressed as follows: 
\begin{eqnarray}
\label{eqn:1}
\frac{d}{dt}P(i)&=&\sum_{j}\left(\Gamma_{ij}P(j)-\Gamma_{ji}P(i)\right),\nonumber\\
\end{eqnarray}
with the normalization condition $\sum_{i}P(i)=1$.  
We solve Eq.~(\ref{eqn:1}) under the stationary condition ($dP(i)/dt=0$), and the probabilities of the spin state, $P(i)$, are obtained. 
$P_Q$ and $P_T$ in the main text correspond to $P(i)$ of $Q(1,2)$ and $T(1,1)$, respectively.

Finally, the current $I$ is described as follows\cite{Bruus}:  
\begin{eqnarray}
I&=&-e\sum_i\left(\sum_{\stackrel{j}{(N_j=N_i+1)}}\Gamma_{ji}^{s}P(i)-\sum_{\stackrel{k}{(N_k=N_i-1)}}\Gamma_{ki}^{s}P(i)\right). 
\end{eqnarray}
In the main text, we define the current flow from the source electrode to drain electrode as positive\cite{Amaha}.


\end{document}